\documentclass[12pt,british]{report}
\usepackage[utf8]{inputenc}
\usepackage[a4paper,twoside,headheight=14.5pt]{geometry}
\usepackage{xcolor}
\usepackage{fancyhdr}
\usepackage{lmodern} 
\usepackage{hyperref} 
\usepackage{listings} 
\usepackage{color}
\usepackage{csquotes}
\usepackage{lastpage}
\usepackage{multicol}
\usepackage{blindtext}
\usepackage{graphicx}
\usepackage{tabularx}
\usepackage[style=numeric,backend=biber,maxnames=2]{biblatex}
\usepackage{mathptmx}
\usepackage{setspace}
\usepackage{textcomp}
\usepackage{diagbox}
\usepackage{multirow}
\usepackage{afterpage}
\usepackage{ragged2e}
\usepackage[british]{babel}

\graphicspath{ {./figures/} }

\usepackage{amsthm}

\begin{document}
\begin{titlepage}

\newcommand{\HRule}{\rule{\linewidth}{0.5mm}} 

\center 


\includegraphics[scale=0.35]{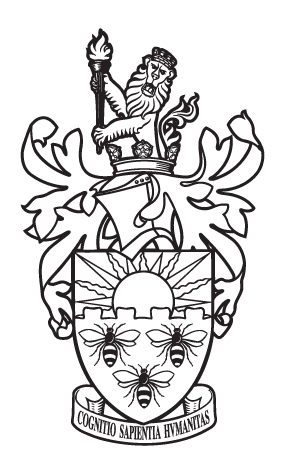}\\[0.5cm] 


\textsc{\Large The University of Manchester}\\[0.2cm] 
\textsc{\large Faculty of Science and Engineering}\\[0.15cm] 
\textsc{\large School of Computer Science}\\[1.5cm] 


{ \huge A Component-Based Approach to Traffic Data Wrangling}\\[1.0cm] 
\textsc{ Third Year Project Report \\submitted in partial fulfilment of the requirements \\for the degree of Bachelor's of Science in\\ Computer Science with Business and Management}\\[3.5cm]

 

\begin{minipage}{0.4\textwidth}
\begin{flushleft} \large
\emph{Author:}\\
Hendrik \textsc{Mölder} 

\end{flushleft}
\end{minipage}
~
\begin{minipage}{0.4\textwidth}
\begin{flushright} \large
\emph{Supervisor:} \\
Dr. Sandra \textsc{Sampaio} 
\end{flushright}
\end{minipage}\\[1.5cm]


\vfill
{\large April 2019}\\[2cm] 
\def\blankpage{%
      \clearpage%
      \thispagestyle{empty}%
      \null%
      \clearpage}
\afterpage{\blankpage}
\end{titlepage}
\onehalfspacing
\tableofcontents
\listoffigures
\listoftables

\addcontentsline{toc}{chapter}{Acknowledgements}
\markboth{\MakeUppercase{Acknowledgements}}{\MakeUppercase{Acknowledgements}}
\chapter*{Acknowledgements}

\begin{quote}
    I would like to thank my supervisor Dr. Sandra Sampaio for her continuous support during all stages of this Third Year Project.
    
    I would also like to thank my family and grandparents, specifically my uncle who provided the financial resources that made these three years in Manchester possible.
    
    Finally, I would like to thank my friends who provided encouragement during the challenging phases of this project.
\end{quote}


\addcontentsline{toc}{chapter}{Abstract}
\markboth{\MakeUppercase{Abstract}}{\MakeUppercase{Abstract}}
\chapter*{Abstract}


We produce an increasing amount of data.
This is positive as it allows us to make better informed decisions if we can base them on a lot of data.
However, in many domains the `raw' data that is produced, is not usable for analysis due to unreadable format, errors, noise, inconsistencies or other factors.
An example of such domain is traffic -- traffic data can be used for impactful decision-making from short-term problems to large-scale infrastructure projects.

We call the process of preparing data for consumption Data Wrangling.
Several data wrangling tools exist that are easy to use and provide general functionality. 
However, no one tool is capable of performing complex domain-specific data wrangling operations.

The author of this project has chosen two popular programming languages for data science -- R and Python -- for implementing traffic data wrangling operators as web services.
These web services expose HTTP (Hypertext Transfer Protocol) REST (Representational State Transfer) APIs (Application Programming Interfaces), which can be used for integrating the services into another system.
As traffic data analysts often lack the necessary programming skills required for working with complex services, an abstraction layer was designed by the author.
In the abstraction layer, the author wrapped the data wrangling services inside Taverna components -- this made the services usable via an easy-to-use GUI (Graphical User Interface) provided by Taverna Workbench, which is a program suitable for carrying out data wrangling tasks.
This also enables reuse of the components in other workflows.

The data wrangling components were tested and validated by using them for two common traffic data wrangling requests.
Datasets from Transport for Greater Manchester (TfGM) and the Met Office were used to carry out the experiments.

\addcontentsline{toc}{chapter}{Copyright}
\markboth{\MakeUppercase{Copyright}}{\MakeUppercase{Copyright}}
\chapter*{Copyright}

\small
\begin{enumerate}
    \item[i) ] The  author  of  this  report  (including  any  appendices  and/or  schedules  to  this  report) owns certain copyright or related rights in it (the “Copyright”) and s/he has given The University of  Manchester   certain   rights   to   use   such   Copyright,   including   for administrative purposes.
    \item[ii) ] Copies  of  this  thesis,  either  in  full  or  in  extracts  and  whether  in  hard  or  electronic copy,  may  be  made only in accordance  with the Copyright, Designs and Patents  Act 1988 (as amended) and regulations issued under it or, where appropriate, in accordance with licensing agreements which the University has from time to time. This page must form part of any such copies made.
    \item[iii) ] The   ownership   of   certain   Copyright,   patents,   designs,   trademarks   and   other intellectual  property  (the  “Intellectual  Property”)  and  any  reproductions  of  copyright works in the thesis, for example graphs and tables (“Reproductions”), which may be described  in  this  thesis,  may  not  be  owned  by  the  author  and  may  be  owned  by  third parties. Such Intellectual  Property  and  Reproductions  cannot  and  must  not  be  made available for use without the prior written  permission of the owner(s) of the relevant Intellectual Property and/or Reproductions.
    \item[iv) ] Further information on the conditions under which disclosure, publication and commercialisation of this thesis, the Copyright and any Intellectual Property and/or Reproductions described in it may take place is available in the University IP Policy (see \url{http://documents.manchester.ac.uk/DocuInfo.aspx?DocID=487}), in any relevant Thesis restriction declarations deposited in the University Library, The University Library’s regulations (see \url{http://www.manchester.ac.uk/library/aboutus/regulations}) and in The University’s policy on Presentation of Theses.
\end{enumerate}
\normalsize

\addcontentsline{toc}{chapter}{Glossary}
\markboth{\MakeUppercase{Glossary}}{\MakeUppercase{Glossary}}
\chapter*{Glossary}

\begin{description}
    \item[API] refers to application programming interface.
    \item[CSV] refers to \textit{comma separated values} file format.
    \item[DW] refers to Data Wrangling.
    \item[DWR] refers to Data Wrangling Request. These are common data analysis problems, which can be answered by performing data wrangling operations on source datasets.
    \item[DS] refers to dataset. We use numbers in subscript to differentiate between datasets.
    \item[GUI] refers to Graphical User Interface.
    \item[HTTP] refers to Hypertext Transfer Protocol.
    \item[OCPU] refers to OpenCPU system.
    \item[REST] refers to Representational State Transfer. It represents constraints used for designing web services -- such services are often called RESTful Web Services (RWS).
    \item[RQ] refers to research question.
    \item[RWS] refers to RESTful Web Services.
    \item[TfGM] refers to Transport for Greater Manchester, the body coordinating public transport in the Greater Manchester area.
\end{description}
\chapter{Introduction}
 
In this chapter we introduce the aims and objectives of this project.
The author defines two research questions that this project will answer in either during the implementation or literature review phase.
The author also defines the scope of the project and provides an overview for the rest of this report.

\section{Motivation}
\label{section:motivation}
It is always necessary to perform Data Wrangling before Data Analysis, as raw data coming from single or multiple sources is often not immediately consumable.
Data wrangling can include steps such as data cleaning, converting data into a consistent format, merging multiple datasets, deriving values from existing values and more.
Data Wrangling is an extremely time-consuming task -- IBM \cite{terrizzano2015data} estimates their data analysts spend around 70\% of their time conducting data wrangling activities as data coming from different sources and in different formats needs to be made easily consumable.
There are two common approaches analysts take when conducting data wrangling activities: (1) developing their own tools using programming languages like R, Python and others or (2) using an existing data wrangling tool that usually includes a convenient graphical user interface (GUI).
However, as \textcite{furche2016data} point out, even these tools often require manual intervention from the analysts.

Traffic data, as seen in the following sections of this report, can often be time series data and processing such data is challenging.
This becomes even more challenging when traffic data needs to be combined with some other data -- it is often hard to draw conclusions purely on traffic data.

\section{Research Questions}
Drawing from the issues described in Section \ref{section:motivation}, we have defined the following research questions (RQ), which will be answered in this report:

\begin{enumerate}
    \item[$RQ_1$.] Is component-based approach appropriate for traffic data wrangling?
    \item[$RQ_2$.] Are current data wrangling tools sufficient for handling complex traffic data wrangling requests?
\end{enumerate}

\section{Project Scope}
In this report, the author aims to answer the research questions by implementing reusable data wrangling components (see Section \ref{section:tavernaComponents}). These components will be used in data wrangling workflows managed and run via a GUI (in Taverna Workbench) to answer two Data Wrangling Requests (DWR).

Manchester traffic data provided by Transport for Greater Manchester (TfGM) APIs \cite{tfgm_apis} is used to validate the data wrangling workflows and components.
Despite the scope of this project being limited to Smart Cities applications, such as analysing traffic data, the author believes the techniques and components from this project are applicable to the wider data science domain.

\section{Project Objectives}
The objectives of this project are to (1) implement reusable data wrangling components, (2) integrate these components into data wrangling workflows which are run and managed via a GUI, and (3) validate the components by running the workflows with data sets provided by TfGM and the Met Office to answer two DWRs.

\section{Report Structure}
First, we give an overview of the importance of data wrangling and various data wrangling tools in Chapter \ref{chapter:background}.
We then proceed to explain the design requirements in Chapter \ref{chapter:design} and define the two DWRs that we will answer in Section \ref{section:dwts}.
In Chapter \ref{chapter:implementation}, we set out the implementation plan and describe the implementation environment. Furthermore, the author provides a summary of aims and conclusions for all three development sprints and an overview of the challenges faced during the implementation phase of the project.
In Chapter \ref{chapter:evaluation} evaluation the data wrangling components is conducted by arranging them into data wrangling workflows.
The author also describes the testing techniques used to validate the components were working as required.
In the final chapter we provide a conclusion and recommendations for further work.

\chapter{Background and Literature Review}
\label{chapter:background}
In this chapter the author has provided an overview of (1) Data Wrangling, including the steps it usually consists of and the most common challenges faced when performing Data Wrangling operations, and (2) different Data Wrangling tools, including the easy-to-use applications with GUIs and programming languages that are used for creating custom Data Wrangling tools. 

\section{Big Data}
According to a report by \textcite{the_economist_2010}, hundreds if not thousands of exabytes ($2^{60}$ bytes) of data are produced every year. 
Scholars such as \textcite{chen2012business} have started to call it the `Big Data era': massive amounts of highly relevant data enables us to make better-informed decisions in many areas \cite{mcafee2012big}.

\textcite{ibm_fourvs} has come up with \textit{Four V's of Big Data} which describe the opportunities and challenges of Big Data: \textit{volume} (enormous amount of data is produced every day), \textit{veracity} (due to the large volume, the correctness of the data can be questionable), \textit{variety} (data comes in different formats; 90\% of the data we produce is unstructured), and \textit{velocity} (data is created at increasing speed).
These four characteristics of Big Data mean that raw data is difficult to consume and analyse.
In order to make it consumable, a set of operations needs to be executed -- this process is called Data Wrangling.



\section{Data Wrangling}
It is very challenging to draw meaningful conclusions based on raw data, because of the \textit{Four V's of Big Data}: volume, veracity, variety, velocity \cite{ibm_fourvs}.
It becomes even more evident when one has to combine multiple sets of raw data while doing data analysis.
\textcite{terrizzano2015data} mentioned that as the number of `raw' data sources grows, then so does the need to curate this data to make it consumable for analysis.
The process of making the raw data consumable means applying a set of data wrangling operations on these data sets.
These operations can include (1) transforming data from one format to another format that is better consumable, (2) cleaning data sets, (3) combining data sets to derive new attributes and more.

\section{Data Wrangling Tools}
On a large scale, data wrangling tools can be categorised into two categories: (1) easy-to-use applications with intuitive GUIs and (2) languages that are used to program custom DW tools. 
Examples of (1) include Open Refine (see Section \ref{section:openrefine}), Trifacta Wrangler (see Section \ref{section:trifacta}), and Taverna Workbench (see Section \ref{section:tavernaWorkbench}). 
Whereas examples of (2) include popular programming languages such as Python (see Section \ref{python}) and R (see Section \ref{r}).
As \textcite{endel2015data} note, although many Data Wrangling tools equipped with a GUI have been developed, in most cases these tools do not cover all aspects of Data Wrangling and are therefore far from complete.

\subsection{OpenRefine}
\label{section:openrefine}
\begin{figure}[htb]
    \centering
    \includegraphics[width=0.9\textwidth]{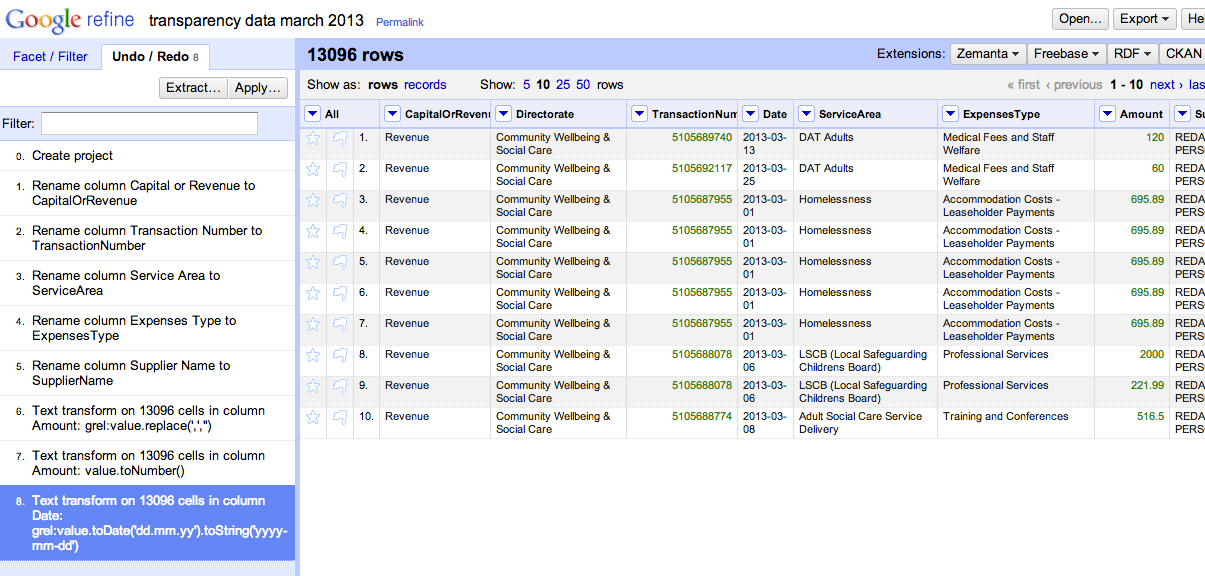}
    \caption{OpenRefine Web-Based Interface}
    \footnotesize{(Source: \fullcite{hirst_2013})}
    \label{fig:openrefine-webbased-interface}
\end{figure}

OpenRefine (also known by its previous name Google Refine) is an open-source data analysis tool originally developed by Google \cite{openrefine}.
It has two interfaces for interaction: an API and a web-based GUI (see Figure \ref{fig:openrefine-webbased-interface}).
Its focus lies in cleaning data by removing inconsistencies and errors from data to improve its quality \cite{groves2016beyond,kusumasari2016data}.
According to its official documentation \cite{willmer_2017} it is also often used for records (or data) normalisation.

A survey carried out by \textcite{magdinier_2018} in 2018, identified the most active users of OpenRefine: librarians and researchers.
The survey also identifies the most common use cases of OpenRefine: normalising data, data transformation from one format to another, and preparing data for another system.
Therefore, we can confidently say OpenRefine is a suitable tool for converting JSON formatted, tree-structured files to a better structured tabular format \cite{permana_2016} -- a task that we need to complete to fulfil the Data Wrangling Request 1 described in Section \ref{section:dwr1}.

However, OpenRefine comes with a few disadvantages.
The most significant being it requires a significant amount of programming skills from its users and has a very steep learning curve \cite{ham2013openrefine}.
Therefore, it might be unusable to anybody without these skills except for performing the most basic Data Wrangling tasks.

\subsection{Trifacta Wrangler}
\label{section:trifacta}
\begin{figure}[htb]
    \centering
    \includegraphics[width=0.9\textwidth]{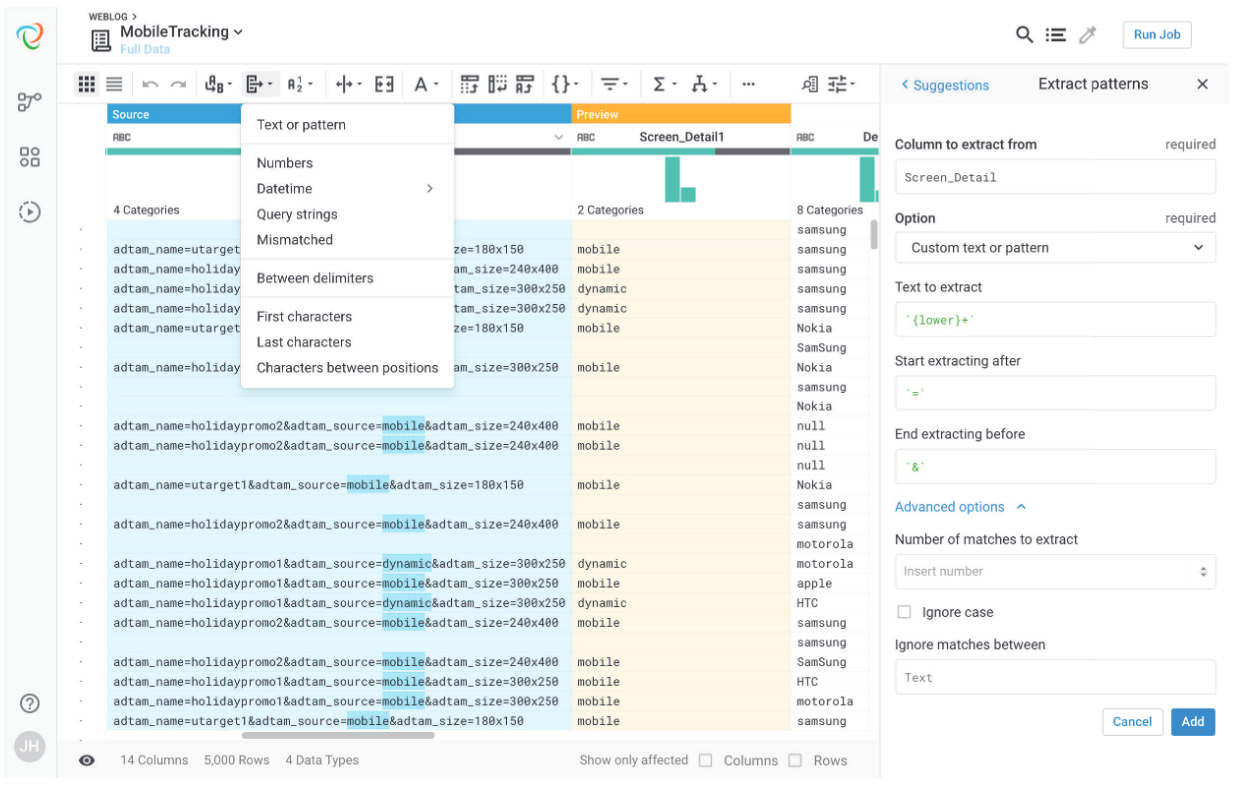}
    \caption{Trifacta Wrangler}
    \label{fig:trifacta}
    \footnotesize{(Source: \fullcite{hellerstein2018self})}
\end{figure}

Trifacta Wrangler (see Figure \ref{fig:trifacta}) is a free data wrangling application \cite{trifacta_2016}.
Trifacta Wrangler performs well for cleaning tabular data \cite{sukhobok2016tabular}.
It also provides functionality for creating data wrangling workflows \cite{simone_2016}.
Its most highlighted functionality includes cleaning datasets, detecting outliers and clustering, and transforming data \cite{hartig2017semantic}.

Despite being a powerful tool for data wrangling, Trifacta Wrangler does not provide an interface for integration with other applications.
As seen from this report, no one tool is powerful enough -- an interface (e.g. HTTP API) for integrating Trifacta with other data wrangling tools would be useful, but is not currently provided.

\subsection{Taverna Workbench}
\label{section:tavernaWorkbench}
\begin{figure}[htb]
    \centering
    \includegraphics[scale=0.3]{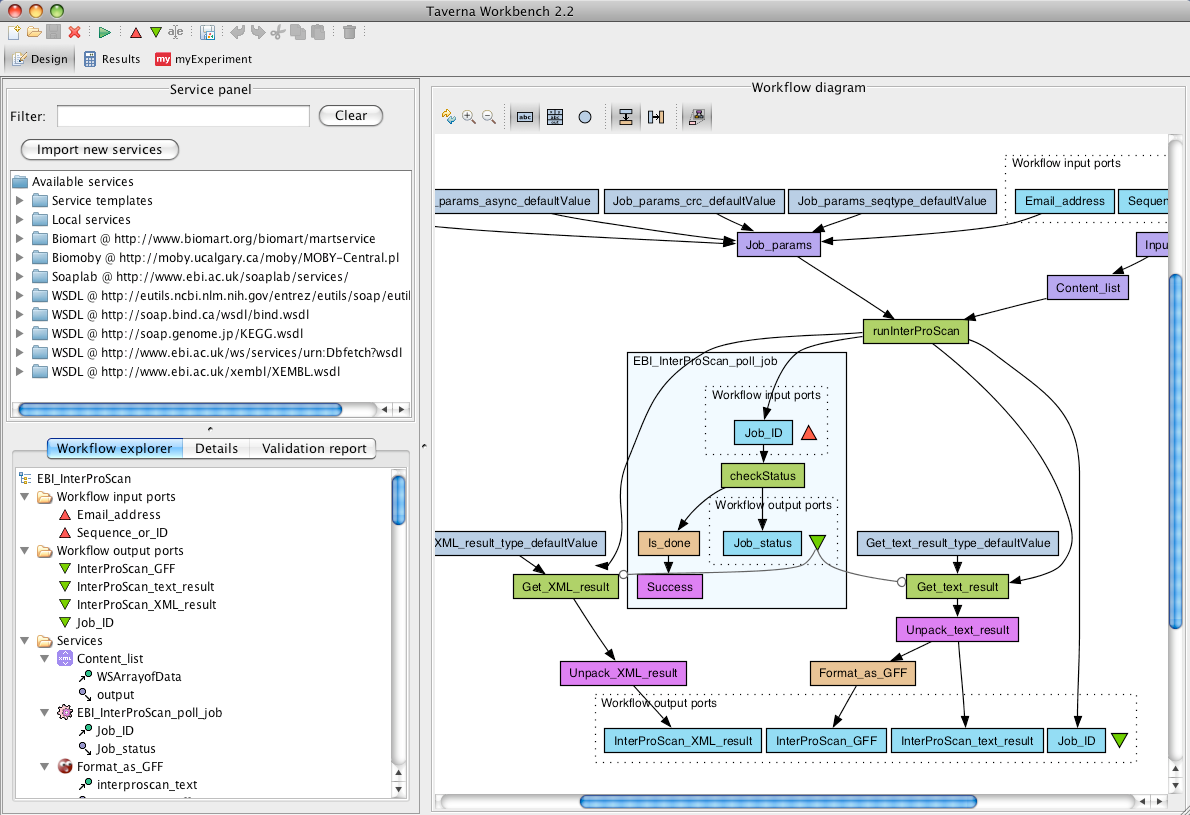}
    \caption{Taverna Workbench}
    \footnotesize{(Source: \fullcite{taverna_intro2019})}
    \label{fig:tavernaScreenshot}
\end{figure}

Taverna Workbench (or `Taverna') is a domain independent environment for executing scientific workflows (see Figure \ref{fig:tavernaScreenshot}) \cite{taverna_intro2019}. 
It enables users to automate tedious and time-consuming manual tasks and has been successfully taken into use in many areas including (but not limited to) bioinformatics to carry out \textit{in silico} experiments. \textcite{oinn2006taverna} have identified three benefits associated with using Taverna: \textit{making tacit procedural knowledge explicit} (workflows are precisely defined and easy to explain), \textit{ease of automation} (automating certain tasks can significantly reduce the time spent on tedious tasks), and \textit{appropriate level of abstraction} (Taverna provides an easy-to-use GUI for interacting with the workflow).

Workflows are the central part of the Taverna program.
Numerous services can be integrated into workflows to perform specific tasks: these can be implemented as Java Beanshell scripts, REST web services, text constants, nested workflows or have another form (see Table \ref{tab:taverna_service_templates} of all service descriptions available in Taverna).
When defined, workflows can be run with specified input (usually data files or text constants) and output ports.
From this point forward, we refer to Taverna Workflows as `workflows' in this report.

Taverna allows parallel execution of independent services in its workflows \cite{wolstencroft_2012}. For example, if two datasets are joined further down in the workflow and require different wrangling operations, these can be run in parallel before merging the two datasets.
In the context of Data Wrangling, workflows can represent Data Wrangling Requests (see Section \ref{section:dwts}), which can be answered by executing the operations specified in the workflow.

\begin{table}[htb]
    \centering
    \caption{Taverna Workbench service templates}
    \label{tab:taverna_service_templates}
    \footnotesize{(Source: \textcite{taverna_core})}
    \normalsize
    \medskip
    \begin{tabularx}{\textwidth}{l | X}
        \hline
        \textbf{Service Type} & \textbf{Description}\\
        \hline\hline
        Beanshell script & A service that allows Beanshell scripts, with dependencies on libraries\\\hline
        Interaction & A service for browser-based interaction with a workflow run\\\hline
        Nested workflow & A service that allows one workflow to be nested within another\\\hline
        REST Service & A generic REST service that can handle all HTTP methods\\\hline
        Rshell & A service that allows the calling of R scripts on R server\\\hline
        SpreadsheetImport & A service that imports data from spreadsheets\\\hline
        Text constant & A string value that one can set\\\hline
        Tool & A service that allows tools to be used as services\\\hline
        xPath Service & A service for point-and-click creation of XPath expressions for XML data\\
        \hline
    \end{tabularx}
\end{table}

\begin{figure}[htb]
    \centering
    \includegraphics[scale=0.6]{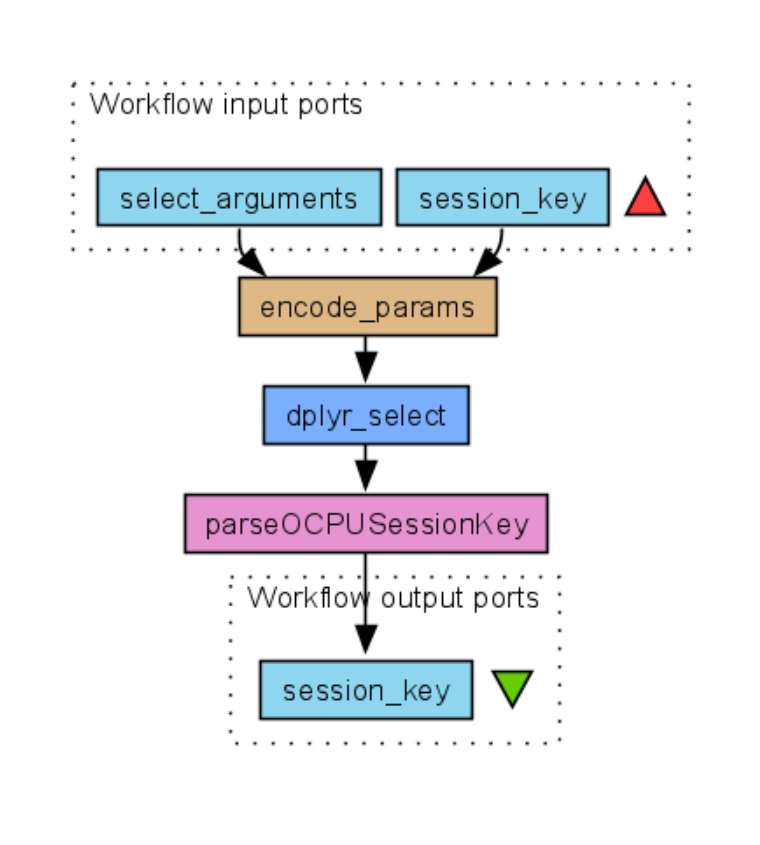}
    \caption{Taverna Component for removing columns from a dataset} \justify\footnotesize{\textit{Workflow input ports} represent the incoming data, that the component will work with. Similarly, \textit{Workflow output ports} represent the data the component will emit -- this can be returned or passed on to other components. Different services are represented in different colours. \texttt{encode\_params} is a Beanshell service that prepares the inputs for the API request carried out by \texttt{dplyr\_select} service. After the REST API request has returned, \texttt{parseOCPUSessionKey} analyses the response and returns the OpenCPU session key (if possible).}
    \label{fig:taverna_remove_columns}
\end{figure}

\label{section:tavernaComponents}
Taverna Components have the same characteristics as `components' in Software Engineering: (1) they are replaceable parts of the workflows and fulfil a clearly defined task, (2) they can be used in all well-structured workflows, and (3) they can communicate with other components via its interfaces (input and output ports) \cite{cai2000component,soiland-reyes_2018}.
For simplification, we can think of Taverna Components as workflows wrapped inside `components' -- with input and output ports -- that perform a clearly defined task in a larger workflow (see Figure \ref{fig:taverna_remove_columns}). 
From here onward, we call Taverna Components `components' in this report.

\subsection{Python}
\label{python}

Python is a programming language with a simple approach to object-oriented programming and high-level data structures \cite{python_software_foundation}.
According to \textcite{vanderplas2016python} it has emerged as the main tool for scientific computer science applications such as data analysis and visualisation during the last decades.
Its dominance is achieved by a large and active ecosystem of packages.
The two most commonly used packages for data manipulation include Pandas for labelled heterogeneous and NumPy for array-based data.

Pandas\footnote{See \url{https://pandas.pydata.org/}.} is an open-source Python library providing tools for high-performance data analysis \cite{pandas}.
It provides an useful \texttt{DataFrame} object, which can be used to manipulate data.
It also has excellent data reading and writing capabilities from and to many different formats including CSV, SQL, Microsoft Excel, text and other formats.

Numpy\footnote{See \url{https://numpy.org}.} is a popular package for scientific computing.
Among other useful features, it includes a robust N-dimensional array object and sophisticated linear algebra functions \cite{numpy}.

Python is also considered a great language web application development \cite{taneja2014python}.
Bottle (sometimes referred to as BottlePy) is an easy-to-use light-weight framework for developing various web applications on Python \cite{summerfield2013python}.
Therefore, if data wrangling operators are implemented in Python, one can easily make these accessible for other tools such as Taverna Workbench via a REST (Representational State Transfer) API (application programming interface) implemented in Python.

\subsection{R}
\label{r}
\begin{figure}[htb]
    \centering
    \includegraphics[width=0.9\textwidth]{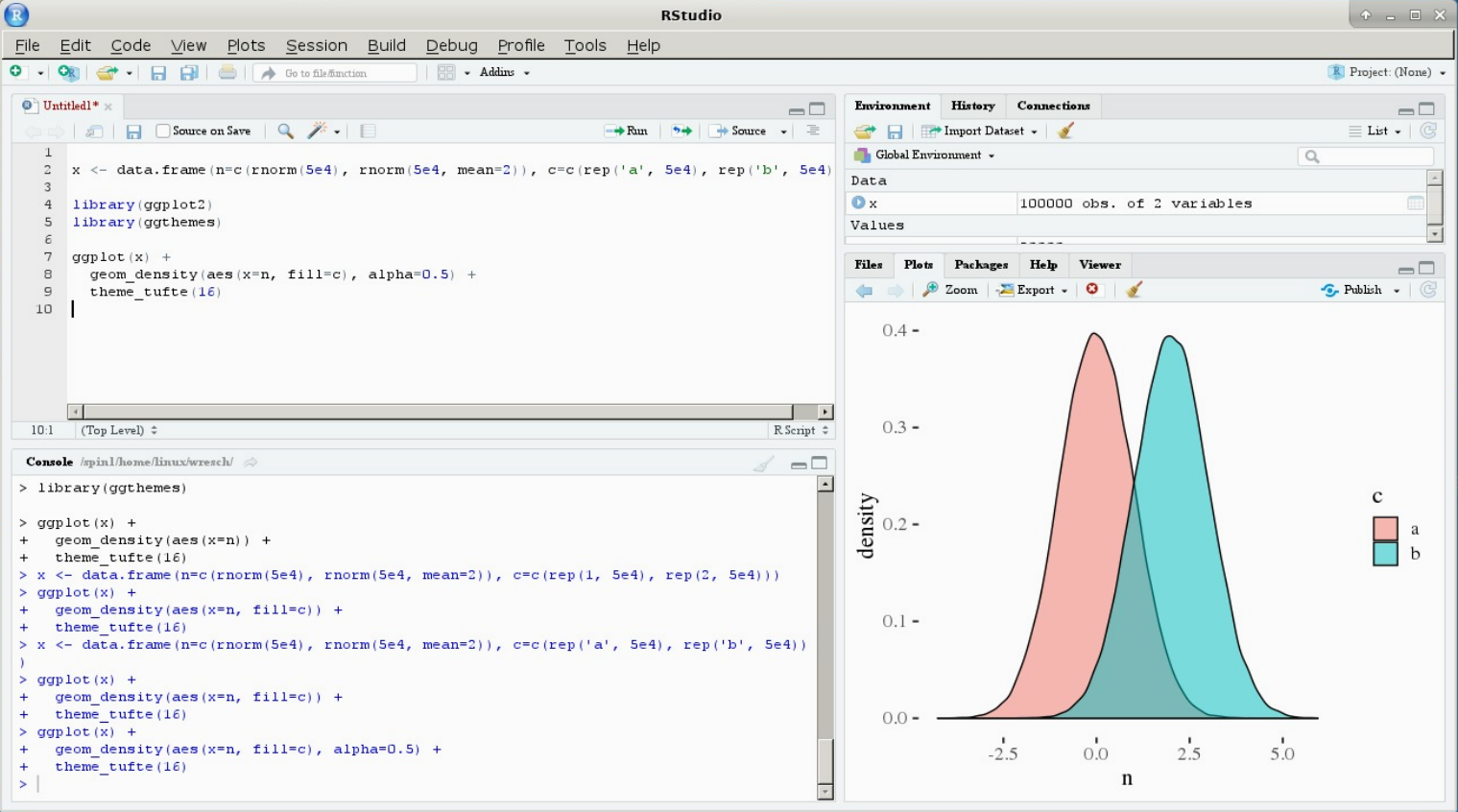}
    \caption{RStudio -- IDE for R}
    \label{fig:rstudio}
    \footnotesize{(Source: \fullcite{nih_rstudio})}
\end{figure}

Like Python, R is a popular programming language that is especially popular among researchers and analysts who handle or process data \cite{tippmann2015programming} -- it is used in a wide range of disciplines, including astrophysics, chemistry, genetics, graphics and others \cite{tippmann_2014}.
R is often used with RStudio\footnote{See \url{http://rstudio.com}.} (see Figure \ref{fig:rstudio})-- an IDE (integrated development environment) for R programming \cite{rstudio_2018}.
It provides a comprehensive overview and is beginner-friendly -- panels for text editor, workspace and history, console, and files, plots, documentation and packages are provided \cite{VanderLoo2012}.

\textcite{wickham_2016} lists the most useful R libraries used for data wrangling: \texttt{tibble} offers additional functionality for R's main data object \texttt{data.frame}, \texttt{readr} provides data importing capabilities for most data formats, \texttt{tidyr} for cleaning data, \texttt{dplyr} for working with relational data, \texttt{stringr} for string manipulation, and \texttt{lubridate} for working with dates and times.
It is clear that working with a programming language requires relevant skills -- however, these libraries significantly simplify this process as they provide functions for a variety of use cases.
Most of these functions are for general use cases.
Hence, if one is not able to generalise the task s/he needs to program, then functions provided by these libraries might not be sufficient and significant programming skills are required.
An example of such case can be working with traffic data.

\subsection{OpenCPU}
OpenCPU (or `OCPU') is a system that combines the functionality of R libraries and HTTP interface to provide an API for scientific work with data.
Specifically, OpenCPU acts as a server and facilitates the mapping between HTTP API requests and R function calls \cite{ooms2014opencpu}.
It stores all function outputs (return values, datasets, matrices, figures) on the server and provides the client a \textit{temporary key} (or `session key') via the HTTP API response.
This is particularly useful when multiple operations need to be carried out on the same data -- instead of uploading data to the server for each function call, the client can refer to the data already present on the server by providing the session key as a function argument.

\section{Summary}
Big Data is changing the way we operate and how we make decisions.
Increasing amounts of data enable us to make better informed decisions in many areas.
However, working with Big Data presents many challenges -- these are summarised by \textit{The Four V's of Big Data}: volume, veracity, variety, and velocity.
In order to effectively use data, it needs to be in shape for analysis -- this is achieved by wrangling the data.

We introduced six data wrangling tools: OpenRefine, Trifacta Wrangler, Taverna Workbench, Python, R, and OpenCPU.
OpenRefine, Trifacta Wrangler, and Taverna Workbench offer easy to use graphical interfaces while R, Python and OpenCPU are programming languages or provide other kind interfaces for communication.
While ease of use is important for data analysts, it often comes with a caveat -- limited functionality or functionality for general use cases only.
Traffic data analysis and wrangling cannot always be generalised enough -- therefore, a significant amount of programming skills are required.
\chapter{Data Wrangling Components Design}
\label{chapter:design}
In this chapter, the author defines the design for the system.
First, the requirements are considered for two data wrangling requests $DWR_1$ and $DWR_2$.
Drawing on these requirements, the author designed data wrangling operations for both requests.
After identifying all data wrangling operators, three Data Wrangling tools were chosen that were the most suitable for implementing these operators.
Finally, an overview of the two-layer system is provided.

\section{Requirements}
\label{section:dwts}

In this report, we use two frequent traffic data analysis requests $DWR_1$ and $DWR_2$ that were constructed by \textcite{sampaio2017} after conducting interviews with traffic data analysts. These requests are described in detail in Sections \ref{section:dwr1} and \ref{section:dwr2}, respectively.
The author has chosen three Data Wrangling tools to be used: Python, R with OpenCPU, and Taverna Workbench.
This selection of tools allows to hide the complexity of the data wrangling tasks behind a layer of abstraction provided by Taverna Workbench while carrying out hard data wrangling problems using components and services implemented in R and Python.

\section{Data Wrangling Request 1 -- $DWR_1$}
\label{section:dwr1}
$DWR_1$ combines three datasets to answer the following Data Wrangling Request:

\begin{quote}
    \textit{What is the typical Friday Journey Time for the fragment of Chester Road stretching from the Poplar Road to the Hulme area between 17:00 and 18:00?} \cite{sampaio2017}
\end{quote}

\begin{figure}[htbp]
    \centering
    \includegraphics[scale=0.6]{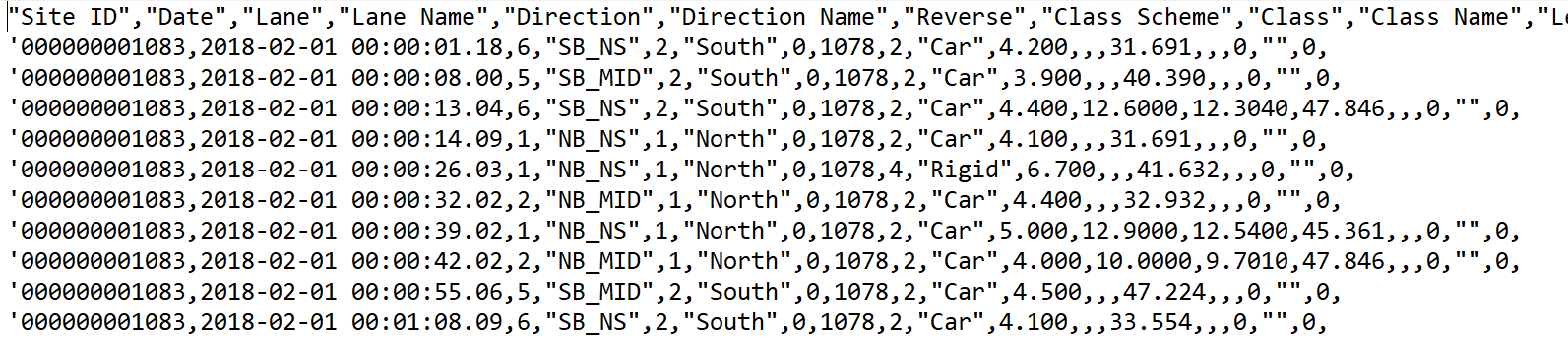}
    \caption{Example excerpt of the structure of $DS1_1$, $DS1_2$, and $DS2_1$}
    \label{fig:traffic_structure_example}
\end{figure}

Two datasets $DS1_1$ and $DS1_2$ (see example structure in Figure \ref{fig:traffic_structure_example}) contain traffic data collected via inductive loops at two locations (sites) in Manchester.
The third dataset $DS1_3$ describes the sites where data for $DS1_1$ and $DS1_2$ were collected and includes the distance between the two sites.
All three datasets $DS1_1$, $DS1_2$, and $DS1_3$ are provided by Transport for Greater Manchester (TfGM) \cite{tfgm_apis}.

A sequence of 11 Data Wrangling tasks is necessary to fulfil $DWR_1$ (see Table \ref{tab:dwr1_operations}). 
Step 3 allows us to reduce the size of $DS1_{12}$ and therefore we are able perform the following operations more efficiently.

\begin{table}[htb]
    \centering
    \caption{Data Wrangling operations for $DWR_1$}
    \label{tab:dwr1_operations}
    \medskip
    \begin{tabularx}{\textwidth}{c | X}
        \hline
        \textbf{Step} & \textbf{Data Wrangling operation} \\\hline\hline
        1 & Perform a join of datasets $DS1_1$ and $DS1_2$. We refer to the joined dataset as $DS1_{12}$.\\\hline
        2 & Upload datasets $DS1_{12}$ and $DS1_3$ to OCPU server.\\\hline
        3 & Remove all irrelevant columns from $DS1_{12}$.\\\hline
        4 & Apply a filter on $DS1_{12}$ to select all Fridays.\\\hline
        5 & In $DS1_{12}$, separate the \texttt{Date} column (includes both, date and time) into two columns: \texttt{Date} and \texttt{Time}.\\\hline
        6 & Clean \texttt{Site.ID} column of $DS1_{12}$ by removing leading zeroes and apostrophes.\\\hline
        7 & Apply a filter on $DS1_{12}$ to only include information within the specified time frame of 17:00 to 18:00 and in the correct directions.\\\hline
        8 & Merge two datasets $DS1_{12}$ and $DS1_3$. We refer to the joined dataset as $DS1_{123}$.\\\hline
        9 & Group entries in $DS1_{123}$ by \texttt{Site.ID} and \texttt{LinkLength}.\\\hline
        10 & Summarise\\\hline
        11 & Calculate journey time.\\\hline
    \end{tabularx}
\end{table}

\section{Data Wrangling Request 2 -- $DWR_2$}
\label{section:dwr2}
The second request, $DWR_2$, combines datasets with different formats and structure to answer the following Data Wrangling Request:

\begin{quote}
    \textit{On rainy weekdays, is the average hourly speed of vehicles on Chester Road, near Poplar Road, lower than that typically observed on dry days?} \cite{sampaio2017}
\end{quote}

\begin{figure}[htbp]
    \centering
    \includegraphics[scale=0.6]{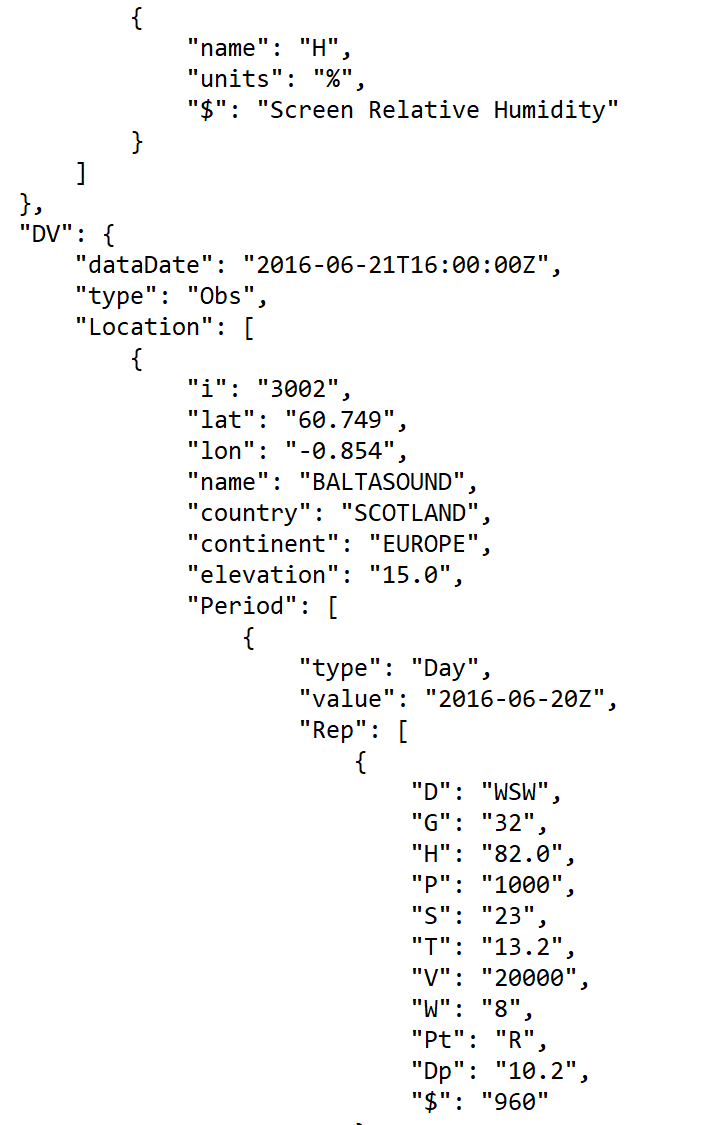}
    \caption{$DS2_3$ in JSON format}
    \label{fig:weather_file}
\end{figure}

Similarly to $DWR_1$, the two datasets $DS2_1$ and $DS2_2$ include traffic data collected via inductive loops and information about the site where the data was collected, respectively.
$DS2_3$ is provided by Met Office and contains weather data of multiple days for thousands of locations in the United Kingdom.
$DS2_3$ is in JSON format and its complex nested structure (see Figure \ref{fig:weather_file}) makes it challenging to convert to a \textit{comma separated values} (CSV) format.
For simplification and considering the weather data we had available, we will only include Fridays in the analysis of $DWR_2$ -- hence we exclude all other weekdays in Step 5 (see Table \ref{tab:dwr2_operations}). 

\begin{table}[htb]
    \centering
    \caption{Data Wrangling tasks for $DWR_2$}
    \label{tab:dwr2_operations}
    \medskip
    \begin{tabularx}{\textwidth}{c | X}
        \hline
        \textbf{Step} & \textbf{Data Wrangling task description} \\\hline\hline
        1 & Make $DS2_1$ and $DS2_2$ available in OCPU.\\\hline
        2 & Remove columns that are not used for analysis from $DS2_1$.\\\hline
        3 & Clean the \texttt{Site.ID} column in $DS2_1$ from leading zeroes and apostrophes.\\\hline
        4 & Merge datasets $DS2_1$ and $DS2_2$. We refer to the merged dataset as $DS2_{12}$.\\\hline
        5 & Apply a filter on $DS2_{12}$ to exclude all days but Fridays.\\\hline
        6 & In $DS1_{12}$, separate the \texttt{Date} column (includes both, date and time) into two columns: \texttt{Date} and \texttt{Time}.\\\hline
        7 & Convert $DS2_3$ from JSON to CSV format. We refer to the converted dataset as $DS2_{3CSV}$\\\hline
        8 & Perform a space-time join operation on datasets $DS2_{12}$ and $DS2_{3CSV}$ -- we refer to the new dataset as $DS2_{123}$\\\hline
        9 & Add \texttt{weatherCond} column to $DS2_{123}$ and populate it with either `wet' or `dry' depending on the weather data.\\\hline
        10 & Group entries in $DS2_{123}$ by \texttt{Site.ID}.\\\hline
        11 & Calculate average speed for each wet and dry day.\\\hline
        12 & Read the necessary data from $DS2_{123}$ and draw a graph for illustrating the (in)difference in average speed for wet and dry days.\\\hline
    \end{tabularx}
\end{table}


\section{Data Wrangling Operators}
After carrying out preliminary experiments to identify the similarities in the Data Wrangling steps of $DWR_1$ and $DWR_2$ described in Tables \ref{tab:dwr1_operations} and \ref{tab:dwr2_operations}, the author has devised 13 generalised Data Wrangling operators that are required.
An overview of these operators is given in Table \ref{tab:dw_operators}.
We carefully analysed the requirements of the operators and compared these to the functionality offered by the Data Wrangling Tools -- the most suitable tools for implementing an operator are marked with `+'.

\begin{table}[htb]
    \centering
    \caption{Data Wrangling Operators}
    \label{tab:dw_operators}
    \medskip
    \begin{tabularx}{0.7\linewidth}{l | c c | c c c}
        \rule{0pt}{1.5cm} Data Wrangling Operator & {\rotatebox[origin=c]{90}{$DWR_1$}} & {\rotatebox[origin=c]{90}{$DWR_2$}} & {\rotatebox[origin=c]{90}{Python}} & {\rotatebox[origin=c]{90}{R \& OCPU}} & {\rotatebox[origin=c]{90}{Beanshell}} \rule{0pt}{1cm}\\[1cm]\hline\hline
        Union files             & + &   & + & + &   \\
        Add and remove columns  & + & + & + & + &   \\
        Filter weekdays         & + & + & + & + &   \\
        Separate date and time  & + & + & + & + &   \\
        Mutate columns          & + & + & + & + &   \\
        Filter function         & + &   & + & + &   \\
        Merge files             & + & + & + & + &   \\
        Group by function       & + & + & + & + &   \\
        Summarise function      & + & + &   & + &   \\
        Perform calculations    & + &   & + & + & + \\
        Transform weather JSON to CSV
                                &   & + & + &   &   \\
        Perform time-space join &   & + & + & + &   \\
        Construct a bar chart   &   & + & + & + &   \\\hline
    \end{tabularx}
\end{table}

\begin{figure}[htb]
    \centering
    \includegraphics[width=0.9\textwidth]{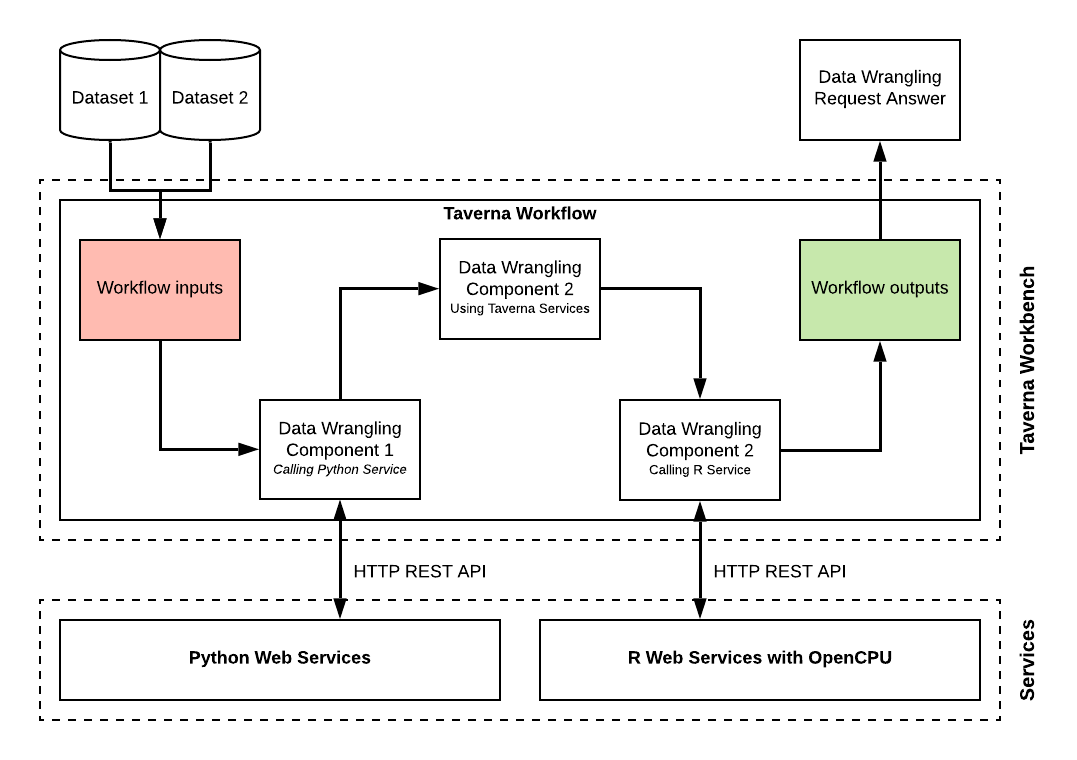}
    \caption{System Design Overview}
    \label{fig:services}
\end{figure}

R and Python are suitable for performing complex traffic data wrangling tasks, whereas Taverna enables to hide the complexity behind a layer of abstraction.
We also used the Beanshell script service provided by Taverna to perform calculations and intermediary operations.
We have described the overall system design on Figure \ref{fig:services}.

Complex data wrangling operators are implemented as Python or R web services (see \textit{Services} on Figure \ref{fig:services}) which are both accessible via HTTP APIs.
To enable easier interaction with the services, we have wrapped these inside Taverna components.
In other words, there are three types of components in the designed system: (1) components that use Python web services in the background, (2) components that use R web services in the background via OpenCPU, and (3) components that use services available in Taverna (described in Table \ref{tab:taverna_service_templates}).
To conclude, we have designed a system where the user is able to execute complex traffic data wrangling functions via Taverna GUI.

\section{Summary}
The author designed a system with two layers: (1) abstraction layer provided by Taverna components and (2) a services layer implemented in R and Python.
Python and R services have been devised from the data wrangling operators required to fulfil $DWR_1$ and $DWR_2$.
These services are wrapped inside Taverna components -- this enables easy interaction with the services and hides the complexity from the user.
Furthermore, the components can easily be reused for other data wrangling requests.
\chapter{Implementation}
\label{chapter:implementation}

The author followed a software engineering model where testing was integrated into implementation sprints -- this is somewhat similar to the agile software engineering principles \cite{beck2001manifesto}.
Data wrangling components were tested in Taverna workflows after every major change to discover any errors and bugs, which were then fixed.
In addition, this approach ensured feedback from user side was taken into account as the author used the same facilities in Taverna to run the workflows as a data analyst would.
The author also used the concept of `sprints' from Scrum \cite{schwaber2010scrum}; implementation phase was split into three software development sprints, which were 1-3 week long periods, each having its aims and goals.

\section{Implementation Plan}
Prior to implementation phase, a plan (see Table \ref{tab:implementation_plans}) was constructed for the whole implementation cycle, which included three sprints.
The author considered the two data wrangling requests $DWR_1$ and $DWR_2$, and the data wrangling operators identified in Table \ref{tab:dw_operators} to set the aims and objectives for each sprint.
Each sprint was set to include approximately the same amount of work.

\begin{table}[htb]
    \centering
    \caption{Implementation Plan}
    \label{tab:implementation_plans}
    \medskip
    \begin{tabularx}{\textwidth}{l | X}
        \textbf{Sprint \#} & \textbf{Aims and Objectives of the Sprint} \\\hline\hline
        Sprint 1 & Implement a Python-based web service for transforming nested, tree-structured weather dataset into a tabular CSV format.\\\hline
        Sprint 2 & Set-up R and OpenCPU, implement the services required by $DWR_1$ and integrate these into Taverna Workbench as components.\\\hline
        Sprint 3 & Implement the remaining services in Python and R that are required for $DWR_2$ and integrate these into Taverna Workbench. Create a comprehensive documentation for the services created in Sprints 1, 2, and 3.\\\hline
    \end{tabularx}
\end{table}

\section{Implementation Environment}
The author has provided a list of packages, tools, and system libraries along with their versions in Table \ref{tab:versions_table}.
Implementation of the system was carried out on Dell XPS 13 machine.
GitHub, a popular service for storing source code, was used for storing source code for the service layer.

\begin{table}[htb]
    \centering
    \caption{Versions of system libraries and tools}
    \label{tab:versions_table}
    \medskip
    \begin{tabular}{c | l  c}
         & \textbf{Description} & \textbf{Version}\\\hline\hline
        \multirow{4}{*}{\rotatebox[origin=c]{90}{System}} & Linux (Ubuntu) & 16.04.5 LTS\\
         & Microsoft Windows & 10 \\
         & Python & 3.5.2\\
         & R & 3.5.2\\\hline
         \multirow{4}{*}{\rotatebox[origin=c]{90}{Software}} & RStudio & 1.1.463\\
        & Taverna Workbench Core & 2.5.0\\
        & MS Visual Studio Code & 1.33.0\\
        & Hyper & 2.1.2\\\hline
        \multirow{7}{*}{\rotatebox[origin=c]{90}{Python Libraries}} & python-dateutil & 2.7.5\\
        & bottle & 0.12.16\\
        & numpy & 1.16.0 \\
        & pandas & 0.23.4\\
        & requests & 2.21.0\\
        & urllib3 & 1.24.1\\
        & atexit & \textit{newest}\\\hline
         
    \end{tabular}
\end{table}

\section{Implementation Sprint 1}
Sprint 1 had one clear goal -- implementation of a Python web service that transforms a complex nested JSON weather dataset into CSV format.
Considering that no existing infrastructure was available, the author also had to implement a Python web server using the BottlePy framework and all intermediary services such as handling data upload and download requests to and from the server.
An overview of the Python web server implementation is provided below.

\begin{itemize}
    \item \texttt{weather\_json2csv} -- this is a service that performs the data transformation from JSON into CSV format.
        This serice was the most challenging part of Sprint 1.
        Preliminary experiments were carried out using a Python programming environment Jupyter Notebook in order to determine the structure of the weather dataset provided by the Met Office.
        Some inconsistencies were found in the dataset structure -- these required handling several special cases in the service implementation.
    \item \texttt{get\_static\_file}\footnotemark -- this service downloads a file specified in the HTTP request from the Python server.
    \item \texttt{download\_from\_path} -- a Python service that parses and saves a file embedded into the request body as text.
    \item \texttt{download\_from\_url} -- this service downloads and saves a file to the server from an URL specified in the request.
        It also checks whether the URL provided points to a downloadable file by checking the \texttt{Content type} value.
        Finally, it fetches the original file name and uses it when saving the file.
    \item \texttt{clear\_downloads}\footnotemark[\value{footnote}] -- this is a server utility function. 
        It deletes all downloaded files before server shutdown is performed.
\end{itemize}

\footnotetext{The author was able to reuse parts of an existing implementation by \textcite{permana_2016}. However, due to differences in Python and library versions a substantial amount of modifications were added. This is noted in the source code in all relevant places.}

\begin{figure}[htb]
    \centering
    \includegraphics[scale=1.2]{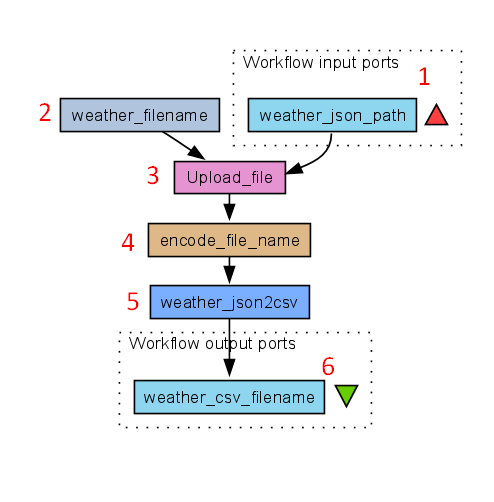}
    \caption{\texttt{weather\_json2csv} Taverna component}
    \label{fig:weatherjson2csv_component}
    \justify\footnotesize{Components are implemented as workflows in Taverna -- hence the naming of input and output ports. (1) represents input dataset, which in our case is the weather file in JSON format. (2) is a text constant which contains a string used for saving the file to the server in the upload service (3). The upload service returns an URL that points to the weather dataset in the Python server. (4) prepares the HTTP request headers and body by including the file URL as a parameter. These request parameters are passed on to (5), which performs the HTTP request to the \texttt{weather\_json2csv} Python service -- when finished it returns an URL to the transformed CSV file and this is used as the output (6) of the component.}
\end{figure}

When the implementation of Python services was completed, the \texttt{weather\_json2csv} service was wrapped inside a Taverna component (see Figure \ref{fig:weatherjson2csv_component}).

\section{Implementation Sprint 2}
Sprint 2 was planned for implementing the data wrangling operators for $DWR_1$.
This included the set-up of R, RStudio, and most importantly - OpenCPU.
In addition to the data wrangling operators necessary for performing $DWR_1$, the author identified some intermediary services and components that are needed -- these include components for uploading and downloading datasets to and from OpenCPU, and handling the temporary key provided by OpenCPU for all datasets.
All services and components implemented during Sprint 2 are summarised below.

\begin{itemize}
    \item \texttt{file\_union} is a Python service used for merging traffic datasets (e.g. $DS1_1$ and $DS1_2$). 
        It is wrapped in a Taverna component that uses file uploading services implemented in Sprint 1.
        The component returns an URL that points to the merged file on the server -- this can be passed on to other components for downloading the file, if needed.
    \item \texttt{upload\_file\_to\_ocpu} is a general component that performs the data upload to OpenCPU server and returns a temporary key, which is a reference to the dataset on the server.
    \item \texttt{remove\_columns}\footnotemark[\value{footnote}] -- this service was implemented in R using the \texttt{dplyr} library's \texttt{select} function. 
        The challenging part of using functions from a library is integrating it into Taverna as a component and passing it the correct arguments in the HTTP request to OpenCPU.
        The formatting and encoding of these arguments can be nontrivial.
    \item \texttt{filter\_weekdays} -- a custom implementation was built using R functions.
        It required filtering out specific weekdays based on date.
    \item \texttt{separate\_datetime}\footnotemark[\value{footnote}] service was used to split the date and time from \texttt{Date} column into two separate columns called \texttt{Date} and \texttt{Hours}. This was implemented in R.
    \item \texttt{clean\_sideid\_column}\footnotemark[\value{footnote}] -- a service for cleaning the \texttt{Site.ID} column of the traffic datasets. This was implemented using \texttt{dplyr} library functions in R.
    \item \texttt{parse\_ocpu\_sessionkey} is a utility service for handling the parsing of OpenCPU temporary key from the HTTP request response.
    \item \texttt{r\_filter} is a general filter function provided by \texttt{dplyr} library in R.
        This was integrated into Taverna in a way which allows the user to specify filter parameters in Taverna.
    \item \texttt{upload\_excel\_to\_ocpu} is another utility function for uploading Microsoft Excel file formats to OpenCPU.
        These file formats require different handling and therefore a separate component was implemented.
    \item \texttt{join\_with\_sitedata}\footnotemark[\value{footnote}] -- this service joins traffic dataset with the traffic site data dataset (e.g. $DS1_3$) in OpenCPU using the standard \texttt{merge} function provided by R.
    \item \texttt{group\_by}\footnotemark[\value{footnote}] service performs a group by operation with the parameters that are specified in Taverna.
    \item \texttt{summarise}\footnotemark[\value{footnote}] service carries out a summarise operation that is provided by the \texttt{dplyr} library in R.
    \item \texttt{extract\_speed\_and\_length} -- this component downloads data from OpenCPU server.
        It works in parallel, downloading speed and distance information from OpenCPU -- this data is necessary for calculating the journey time, which is the final task of $DWR_1$.
        This component used multiple R services, including \texttt{summarise} and \texttt{subset} functions from \texttt{dplyr} library.
    \item \texttt{calculate\_journey\_time} carries out the final task of $DWR_1$, the calculation of journey time using data downloaded from OpenCPU.
        This service was implemented as a Beanshell script inside Taverna.
\end{itemize}

\section{Implementation Sprint 3}
The final implementation sprint 3 was planned for implementing the remaining services for $DWR_2$ -- these were services that were not used for $DWR_1$.
This sprint also included writing a comprehensive documentation for all services implemented in Sprints 1, 2 and 3, and creating a set-up guide for OpenCPU server and R, which proved to be challenging without any prior experience.
A list of services and documentations is noted below.

\begin{itemize}
    \item \texttt{time\_space\_join}\footnotemark[\value{footnote}] is a service implemented in Python. It joins the weather file to the traffic data file by time and space (distance) with predefined buffers. 
        For example, most likely the weather data measured at a mile distance of the traffic sensor location is relevant for the traffic data around that time period, so this information will be merged.
    \item \texttt{add\_weather\_columns} service was implemented in R. 
        It added a new column \texttt{weatherCond} to the data file with a value of `dry' or `wet' depending on the weather information that had been merged with the file.
    \item \texttt{create\_graph}\footnotemark[\value{footnote}] is a custom function implemented in R that constructed a bar chart for average speeds on dry and wet days for $DWR_2$.
    \item \textit{Wrangling Services REST API Documentation} is a documentation for Python web services (see Appendix \ref{chapter:appendix_a}).
        It included a HTTP request template, return types, request parameters, and examples for every service.
        OpenCPU automatically generates documentation as HTML pages for the services it has available.
    \item \textit{Set-Up Guide for OpenCPU} is a document containing instructions for setting up R and OpenCPU, installing and creating R packages, and exposing these via OpenCPU server.
\end{itemize}

\section{Implementation Challenges}
The author faced several challenges during the three implementation sprints.
During sprint 1, it became evident that due to inconsistencies in the JSON weather file provided by the Met Office, many special cases needed to be handled.
These inconsistencies were identified by carrying out experiments with the Jupyter Notebook software.

There were two very challenging parts in sprint 2.
Setting up OpenCPU server with R was found out to be more complex than initially expected.
OpenCPU documentation did not provide any examples for setting up new R packages in the server.
Once the OpenCPU was set up, it was found out that some functions from \texttt{dplyr} and other libraries do not take function arguments in the same format via OpenCPU as they do in RStudio.
Further research into R library functions and OpenCPU was carried out to successfully wrap these services inside Taverna components.

It is also important to note the author did not have prior experience with R and Python languages and libraries.
The author used books and various other resources to learn the basic concepts of both languages needed for completing the implementation phase of this project.
The documentation written during sprint 3 could potentially avoid running into some of these challenges in the future.

\section{Summary}
The author used methods from two software engineering approaches -- Agile and Scrum development.
The implementation phase of the project was split into three implementation sprints, which all had their own aims and objectives.
Each sprint was 1-3 weeks in length and included iterations of development and testing.

To summarise, sprint 1 was planned for implementing a service for transforming weather dataset into tabular (e.g. CSV) format.
Sprint 2 concerned implementing services required by $DWR_1$.
During sprint 3, the author implemented the necessary services for $DWR_2$ and wrote documentation for services implemented in all three sprints. 

The author faced many challenges during the three implementation sprints.
However, these were overcome by carrying out further experiments and research.
\chapter{Testing and Evaluation}
\label{chapter:evaluation}

Testing plays an important role in software development often taking up more than 50 per cent of the total development time \cite{myers2004art}.
In this project, the author's main approach to testing was integration testing, more specifically incremental integration testing.
Integration testing focuses on relations and interfaces between different parts of a system \cite{linnenkugel1990test}.

In the context of data wrangling workflows, this approach is used to test the interactions between the data wrangling components.
As common to incremental testing, components were gradually added to the workflows while carrying out checks and tests in between to validate the correct operation of the components and the underlying Python and R services.

\section{Evaluation Criteria}

For evaluating the data wrangling components (and the Python and R services embedded inside them), the author arranged these components into Taverna workflows representing the two data wrangling requests $DWR_1$ and $DWR_2$.
These workflows were executed with the example data sets provided by TfGM and the Met Office.
The successful execution of the workflows would confirm the components are working and communicating with each other in a specified manner.

\section{Evaluation of $DWR_1$}
\begin{figure}[htbp]
    \centering
    \includegraphics[scale=0.9]{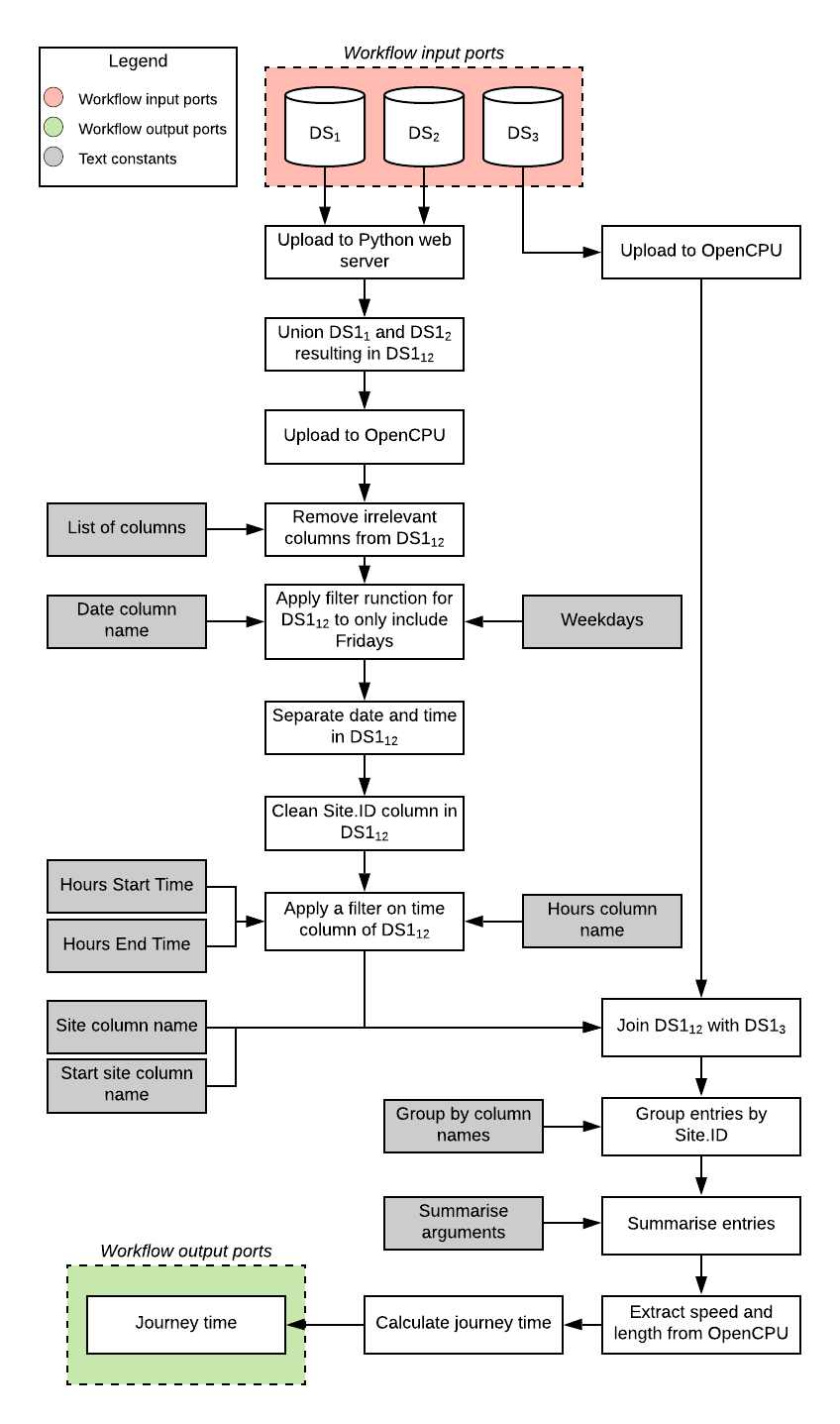}
    \caption{$DWR_1$ Workflow Summary}
    \label{fig:dwr1_workflow}
\end{figure}

A Taverna workflow was constructed based on the requirements of $DWR_1$. 
Data wrangling components were used to construct a workflow corresponding to the requirements of the data wrangling request. 
This workflow is summarised in Figure \ref{fig:dwr1_workflow}.

Incremental integration testing was carried out throughout the process.
Minor implementation bugs were identified during the testing of $DWR_1$ workflow -- this extended the planned duration of implementation sprint 2.
However, it is important to note by the end of sprint 2, these bugs were fixed and the workflow as a whole performed as expected by returning a valid journey time for datasets $DS1_1$, $DS1_2$, and $DS1_3$.

\section{Evaluation of $DWR_2$}
\begin{figure}[htbp]
    \centering
    \includegraphics[scale=0.9]{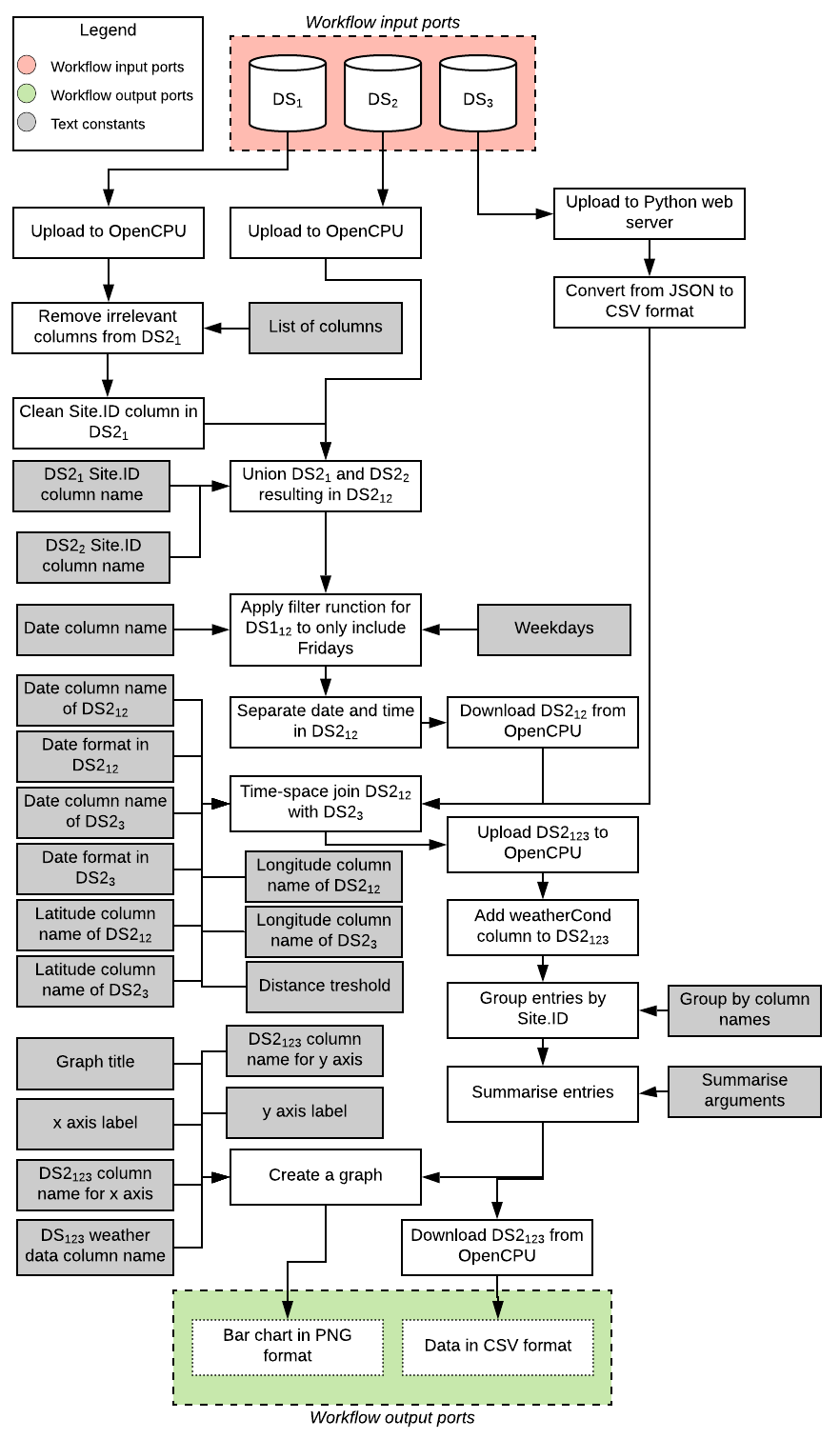}
    \caption{$DWR_2$ Workflow Summary}
    \label{fig:dwr2_workflow}
\end{figure}

A data wrangling workflow was also implemented for the second data wrangling request $DWR_2$.
Components from spints 1, 2, and 3 were used to construct the workflow (see Figure \ref{fig:dwr2_workflow}).
The successful execution of this workflow helped validate all components performed as expected.

Similarly to the evaluation of $DWR_1$, the author found minor issues with the implementation of components used for $DWR_2$ workflow when carrying out incremental integration testing.
Most of these bugs were related to the way the author had wrapped the Python and R services inside Taverna components.
These issues were fixed by the end of final implementation sprint.

\section{Summary}
Concepts of integration and incremental testing were used by the author to validate the implementations of all data wrangling components.
This testing was carried out by evaluating the execution of the components that were arranged into data wrangling workflows corresponding to $DWR_1$ and $DWR_2$.

Minor implementation bugs were discovered in the testing phase -- these were fixed during the implementation sprints 2 and 3.
The majority of the discovered bugs concerned correctly wrapping Python and R web services inside Taverna components.

\chapter{Conclusions and Future Work}

In this chapter the author summarises the work completed for this project.
Limitations of the work and future work possibilities will also be described.

\section{Conclusions}
The aim of this project was to compare different data wrangling tools for traffic data wrangling purposes.
A component-based approach was picked by the author and a system with two layers was designed.
The first, services layer included data wrangling operators implemented in R and Python.
These services were implemented as web services and later wrapped inside Taverna components as part of the second, abstraction layer of the system.

The author identified that existing data wrangling tools are good for general purposes.
However, complex and domain-specific data wrangling tasks require more advanced tooling.
The most popular advanced data science tools are two programming languages -- R and Python.
These languages have a steep learning curve and are note usable by users who do not have programming experience.

The author implemented some complex and some general data wrangling operators in R and Python.
These were implemented as web services and were later wrapped inside Taverna components.
This offered two advantages: (1) Taverna Workbench offers an intuitive GUI for easy user interaction, components can be arranged into a sequence by dragging and dropping them in place, and (2) components become easily reusable -- they can be easily used for new data wrangling requests.

\section{Limitations}
Most of the data wrangling components were implemented in a way that enables the reuse of these components for new workflows in Taverna.
However, some components serve very specific traffic data wrangling tasks -- reusing such components might require some additional changes to the components themselves or to the underlying services in R and Python.

Another limitation of this approach is that Taverna components are not usable outside Taverna Workbench application (i.e. in other data wrangling tools).
As the web services expose a HTTP API, they can be integrated into other systems but this process might require programming competence.

Finally, it is important to mention, that due to the design of this two-layer system, Python and OpenCPU servers must be running to enable the use of the created data wrangling components.

\section{Skills Acquired}
The author acquired many new skills throughout different phases of this project.
The most noticeable are listed below.

\begin{itemize}
    \item The author did not have prior experience in working with Big Data.
    \item The author had no prior experience in programming with R.
        R proved to have a steep learning curve due to the considerable differences in its syntax compared to other programming languages.
        The author learned the basics of programming in R and RStudio, including how to effectively use data science libraries in R.
    \item The author lacked previous programming experience in Python.
        The easily readable syntax of Python helped to quickly learn the basics of Python programming.
        However, handling HTTP requests and operating with large files imposed some challenges.
\end{itemize}

\section{Further Work}
The author believes further work can be carried out drawing from this project.
Any future work has a great potential to automate or further simplify the tasks of traffic data analysts who operate with complex traffic data.

Firstly, library of data wrangling components should be created.
myExperiment\footnote{See \url{https://www.myexperiment.org}.}\cite{myexperiment} offers such functionality.
It would be useful to publish the existing components and any further components to either myExperiment or a similar service. 

Secondly, more general data wrangling services should implemented.
This would allow the creation of more complex, domain-specific data wrangling operators, which can be constructed from the components providing general functionality.

\clearpage
\onehalfspacing
\setlength{\parskip}{1em}
\addcontentsline{toc}{chapter}{Bibliography}
\markboth{\MakeUppercase{Bibliography}}{\MakeUppercase{Bibliography}}
\printbibliography

\appendix
\chapter{Python Services Documentation Example}
\label{chapter:appendix_a}

\begin{figure}[htbp]
    \centering
    \includegraphics[width=\textwidth]{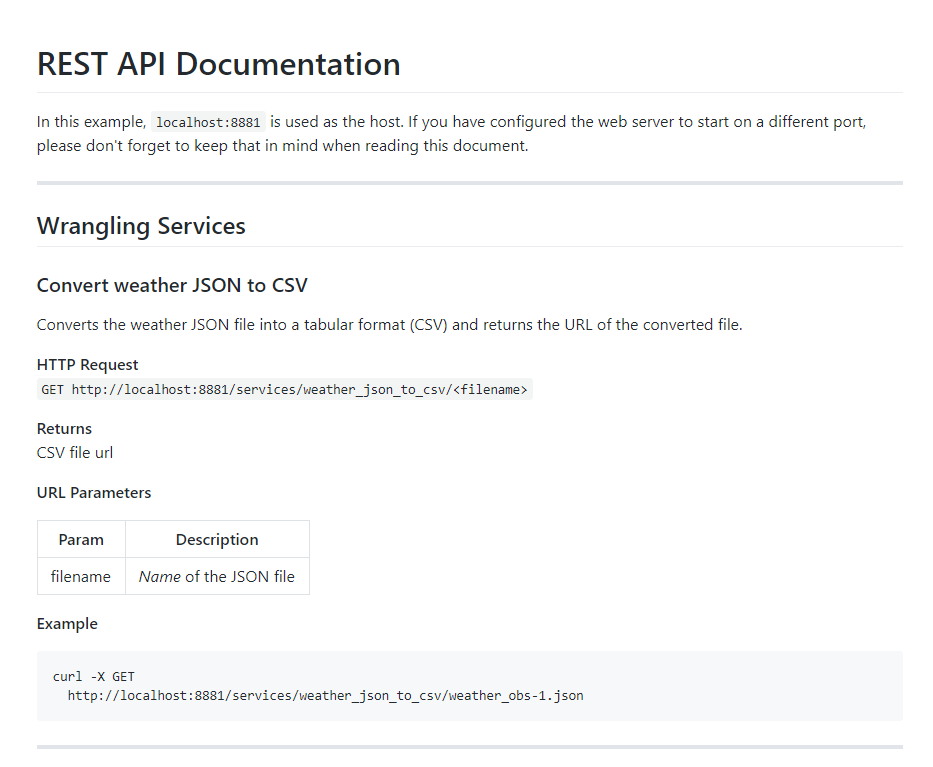}
    \caption{Example of the API Documentation for Python services}
    \label{fig:python_documentation}
\end{figure}
\end{document}